\documentclass[prl,twocolumn,final,superscriptaddress,showpacs,a4paper]{revtex4}

\usepackage{graphicx} 
\usepackage{dcolumn} 
\usepackage{bm} 
\usepackage[T1]{fontenc} 

\begin{document}

\title{Formation of spatial shell structure in the superfluid to Mott insulator transition}

\author{Simon F\"olling}\email[Electronic address: ]{foelling@uni-mainz.de}
\author{Artur Widera}
\author{Torben M{\"u}ller}
\affiliation{Johannes Gutenberg-Universit{\"a}t, Staudingerweg 7, 55118 
Mainz, Germany}
\author{Fabrice Gerbier}
\affiliation{Laboratoire Kastler Brossel, Ecole Normale Sup\'erieure, 24 rue Lhomond, 75005 Paris, France}
\author{Immanuel Bloch}
\affiliation{Johannes Gutenberg-Universit{\"a}t, Staudingerweg 7, 55118 
Mainz, Germany}

\date{\today}

\begin{abstract}
We report on the direct observation of the 
transition from a compressible superfluid to an incompressible Mott insulator by recording the in-trap density distribution of a Bosonic quantum gas in an optical lattice.
Using spatially selective microwave transitions and spin changing collisions, we are able to locally modify the spin state of the trapped quantum gas and record the spatial distribution of lattice sites with different filling factors. As the system evolves from a superfluid to a Mott insulator, we observe the formation of a distinct shell structure, in good agreement with theory. 
\end{abstract}


\maketitle

Ultracold atoms in periodic potentials have proven to be highly tunable model systems for investigating fundamental problems of condensed matter physics \cite{jaksch05a,bloch05a,morsch06}. 
Among those, the transition from a superfluid (SF) phase to a Mott
insulating (MI) phase has attracted much attention (see
\cite{jaksch05a,bloch05a} for recent reviews and references). So
far, experiments have relied mostly on the time-of-flight technique to
probe the many-body properties of the gas cloud. Unfortunately, one
of the most important characteristics of a MI state, its
incompressibility, remains hidden in such measurements. In an
overall confining potential, as is always present for trapped gases in a lattice potential, incompressibility results in the formation of a
shell structure. This structure consists of a succession of large concentric ``Mott plateaus'', where the on-site occupation is pinned to integer values with suppressed
fluctuations and abrupt interfaces between two of these shells
\cite{jaksch98,kashurnikov02,batrouni02,wessel04,bergkvist04,pollet04,demarco05a,rey06a}.
In contrast, the condensate density distribution in the weakly
interacting regime mirrors the smooth shape of the confining potential
\cite{stringaribook}. This dramatic change has so far eluded
direct observation \footnote{The observation of such spatial shell
structures also critically depends on the temperature of the system, since this structure does not survive temperatures larger than $k_B T \approx 0.2\,U$ \cite{demarco05a,schmidt06a,gerbierinprep}.}, although earlier experiments support the existence of a shell structure
\cite{gerbier05a,sengupta05,gerbier06a}.

\begin{figure}
\includegraphics[scale=1.05]{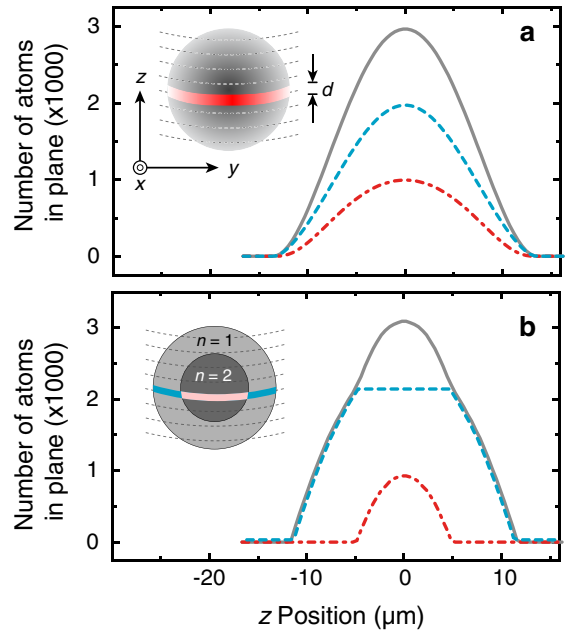}
\caption{Density distribution of a superfluid {\bfseries (a)} and a Mott insulating state {\bfseries (b)} in a lattice with harmonic confinement. Grey solid lines denote the total density profiles, blue  (red) lines the density profiles from singly (doubly) occupied sites. A vertical magnetic field gradient is applied which creates almost horizontal surfaces of equal Zeeman shift over the cloud (dashed lines in inset). A slice of atoms can be transferred to a different hyperfine state by using microwave radiation only resonant on one specific surface (Coloured areas in insets). Spin changing collisions can then be used to separate singly (blue) and doubly occupied sites (red) in that plane into different hyperfine states.
}
\label{fig:schematic}
\end{figure}

Here we report on a direct observation of the in-trap density distribution with a high spatial resolution. This is achieved by locally modifying the spin state using a technique similar to magnetic resonance imaging and to the RF addressing technique in \cite{ott04b}. Spin changing collisions \cite{widera05} allow us to independently measure the integrated density profiles both of the total atom distribution and of sites with a specific occupation number. From such integrated profiles we obtain clear signatures of the emergence of Mott plateaus with uniform occupation above the MI transition. The measured plateau radii agree well with a simple model assuming a fully incompressible system, zero temperature and zero tunnelling.

Our experiment can be outlined as follows: After preparing the gas in the optical lattice at the desired depth, the depth is increased rapidly in order to freeze out the spatial distribution in the trap. The atom cloud is situated inside a magnetic field with a vertical gradient and correspondingly almost horizontal surfaces of equal field strength as depicted in Fig.~\ref{fig:schematic}. A microwave pulse on a field-sensitive transition can now selectively address atoms from only a slice of the cloud where the transition is shifted into resonance by the field. Atoms are transferred to a different hyperfine state within the spectral width of this pulse, corresponding to a region with a thickness $d$ of typically a few micrometers. This allows the determination of the fraction of atoms in the slice using spin state-selective detection.
Repeating the experiment for different microwave frequencies yields a complete profile of the atom density along the direction of the field gradient (Grey curves in Fig.~\ref{fig:schematic}).

In order to distinguish doubly occupied sites from others, we transfer the atoms of the slice to a hyperfine state in which they can undergo spin-changing collisions. A controlled spin oscillation then allows atoms on the doubly occupied sites to make a transition to another, previously unoccupied, spin state (red areas in the inset in Fig.~\ref{fig:schematic}b), not affecting sites with occupation numbers other than $n=2$ (see \cite{gerbier05c} for details). By subsequently measuring the population in each of the states, we obtain the spatial distribution of the overall density and of the density of doubly-occupied sites independently.

To discuss the expected shapes and distinct features of the resulting integrated density profiles, let us first consider the weakly interacting regime of a BEC.
In this case, the full 3D distribution of the mean atom number across the lattice is the Thomas-Fermi distribution $n_{\text{TF}} = n_0~\text{max}(0,1-(x^2+y^2+z^2)/R_{\text{TF}}^2)$, where $R_{\text{TF}}$ is the Thomas-Fermi radius and $n_0$ the central 3D density. The integrated profiles are obtained through integration over the $x-$ and $y-$axes and a slice of thickness $d$ on the $z-$axis (see Fig. 1). The integrated distribution of sites occupied by a specific number of atoms can be readily obtained from the 3D mean atom number distribution, by assuming a Poissonian atom number statistics on each site before carrying out the integration over the slice. In this limit, it is found that the distributions of both doubly and singly occupied sites have a smooth shape very similar to the profile of the overall density, and with a similar width (see Fig.~\ref{fig:schematic}a). Doubly occupied sites can in fact be found at any place inside the cloud, all the way out to the border.     

In contrast, the 3D density distribution of a Mott insulator state in the low atom number limit for given trap parameters is expected to have a spherical shape, with constant unity filling up to a radius $R$. At this radius, the density abruptly drops to zero when assuming negligible tunneling and zero temperature. The integrated density profile of such a sphere with constant density is given by an inverted parabola of the form $\nu(z,R)=8 \pi \lambda_{\text{lat}}^{-3} \cdot d  \cdot \text{max}(0,R^2-z^2)$, where $\lambda_{\text{lat}}$ is the lattice wavelength. The radius of this parabola $R$ is equivalent to the radius of the spherical density distribution in 3D, and the distribution has sharp edges at this distance from the center. 
For higher atom numbers, a two-shell system is expected to develop, with an inner spherical core of doubly occupied sites and radius $R_2$ and an outer shell with occupation number $n=1$ of radius $R_1$. Here, the integrated density profile of the $n=2$ core is given by a parabola $\nu_2(z)= 2 \nu(z,R_2)$ (dash dotted curve in Fig. 1b). The corresponding $n=1$ shell is hollow, and leads to an integrated distribution $\nu_1(z) = \nu(z,R_1)-\nu(z,R_2)$. This is a truncated flat top parabola (dashed curve in figure 1b), in stark contrast to what is observed in the weakly interacting case. Note that in the MI case also the radii of the integrated profile corresponding to singly occupied sites and the one corresponding to doubly occupied sites are markedly different.  

\begin{figure}
\includegraphics[scale=0.9]{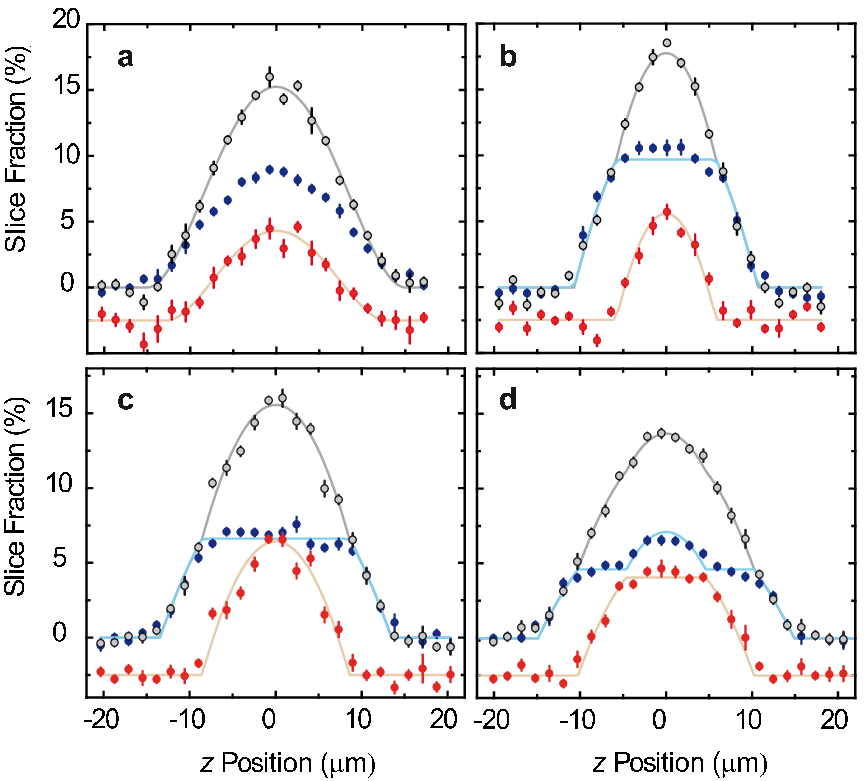}
\caption{Integrated in-trap density profiles of the atom cloud for different lattice depths and atom numbers: {\bfseries (a)} $1.0\times 10^5$ atoms in the superfluid regime ($V_0=3 E_{\text{r}}$), {\bfseries (b)} $1.0\times 10^5$ atoms in the Mott regime ($V_0=22 E_{\text{r}}$), {\bfseries (c)} $2.0\times 10^5$ atoms, {\bfseries (d)} $3.5\times 10^5$ atoms. The grey data points denote the total density distribution and the red points the distribution of doubly occupied sites. The blue points show the distribution of sites with occupations other than $n=2$. The solid lines are fits to an integrated Thomas-Fermi distribution in {\bfseries (a)}, and an integrated shell distribution for {\bfseries (b)} to {\bfseries (d)}. The $n=2$ data points are offset vertically for clarity.}

\label{fig:profiles}
\end{figure}


We prepare our atomic sample by loading a quasipure BEC with up to $5\times 10^5$ $^{87}$Rb atoms prepared in the $\left|F=1, m_F=-1\right\rangle$ state in a QUIC magnetic trap \cite{esslinger98a}. Combined with a homogeneous offset field, this results in an almost isotropic magnetic trap with a frequency $\omega_T\approx2\pi\times15\,$Hz for the three axes. 
Because of gravity, the atoms are displaced from the magnetic field minimum, leaving them in a gradient field of ${\partial B} / {\partial z} = 3.4\,\text{G}/\text{mm}$ which provides the spatial selectivity of the microwave pulse. 
The curvature of the surfaces of equal magnetic field due to the lateral confinement of the trap has no significant effect on the integrated density profiles within our measurement accuracy.
The condensate is loaded into an optical lattice created by mutually orthogonal standing waves with a wavelength of $\lambda_{\text{lat}}=843\,$nm and an average waist of $144$\,$\mu$m by ramping up the potential depth to a final value $V_0$ (typically $22\,E_{\text{r}}$) within $160\,$ms \cite{gericke06a}. In this configuration the external confinement is provided by a combination of the magnetic trap and the Gaussian profile of the lattice beams. For analyzing the density distribution, the lattice depth is then rapidly increased to $40\,E_{\text{r}}$ within $100$\,$\mu$s in order to freeze out the distribution. Here, $E_{\text{r}}=h^2/{2 m \lambda_{\text{lat}}^2}$ is the recoil energy of the lattice light, with $m$ the mass of a single atom.

A microwave $\pi$-pulse of 150\,$\mu$s length is then used to transfer one slice of atoms from $\left|F=1, m_F=-1\right\rangle$ to $\left|F=2, m_F=0\right\rangle$. A second $\pi$-pulse is applied to bring all atoms from $\left|F=2, m_F=0\right\rangle$ to $\left|F=1, m_F=0\right\rangle$ within $5.6$\,$\mu$s. 
The magnetic gradient field is then switched off in favor of a homogeneous offset field of $1.2$~G. At this magnetic  field, spin-changing collisions within the $F=1$ manifold are strongly suppressed, even on lattice sites which contain more than one atom \cite{widera05}. 
Using microwave dressing \cite{gerbier05c}, the collisional spin dynamics can be fully controlled and doubly occupied sites selectively transferred from the $\left|m_F=0, m_F=0\right\rangle$ two-particle state to the $\left|m_F=-1, m_F=+1\right\rangle$ state. Finally the optical lattice is switched off and, during a time-of-flight period of 13~ms, the spin states are separated by a 3.5~ms Stern-Gerlach gradient field pulse. The separated atom clouds are imaged using standard absorption imaging technique.


In order to obtain the integrated density profiles from the resulting set of images, the atoms corresponding to the three Zeeman sublevels are counted by integrating over the cloud areas of the absorption images, yielding the population numbers $N_{m_F=-1}$, $N_{m_F=0}$ and $N_{m_F=+1}$ for the three $m_F$-states, respectively. Atoms from outside the slice are in the $m_F=-1$ Zeeman sublevel, atoms from inside the slice which were not in doubly occupied sites are in the $m_F=0$ sublevel.
Due to the spectral profile of the slicing pulse, the transfer of both atoms of a pair has an overall efficiency of $\alpha=0.58$ across the slice. After the collisional interaction, one atom from each transferred pair is in the $m_F=+1$ state and the other in the $m_F=-1$ state. 
The number of atoms among those in the slice which originated from doubly occupied sites is therefore $\nu_2=2 \cdot N_{m_F=+1}/\alpha$. The total number of atoms addressed by the slicing pulse is $\nu_{\text{slice}}=N_{m_F=0}+2\cdot N_{m_F=+1}$ and $N_{\text{total}}$ the overall number of atoms in that image.

Some resulting density profiles can be seen in Figure~\ref{fig:profiles}. Using the known field gradient and magnetic moment, we translate microwave frequency into position. 
We plot the fraction of the atoms in the slice $\nu_{\text{slice}}/N_{\text{total}}$ as well as the fraction from doubly occupied sites, $\nu_2/N_{\text{total}}$ and their difference. This difference corresponds to the number of atoms from sites with occupation numbers other than $n=2$ (mostly 1 and 3 for our parameters). 
In the weakly interacting state ($V_0=3~E_{\text{r}}$), the overall profile is fitted well by an integrated Thomas-Fermi profile with central density and outer radius as fit parameters. The same function was fitted to the profile of the doubly occupied sites, leading to a very similar radius (see Fig. \ref{fig:radiiratio}).

For deeper lattices ($V_0 = 22 E_{\text{r}}$), the measured integrated density distribution is strikingly different from the one in the weakly interacting regime. For an overall atom number of $2\times 10^5$ atoms, we observe that the distribution of doubly occupied sites is contained in a much smaller radius. Moreover, the shape of the number distribution is that of an inverted parabola with sharp edges instead of the smoother one found for low lattice depths. The difference signal between the profiles for the integrated overall atom number distribution and the $n=2$ distribution exhibits a truncated parabola profile, as is expected for a 3D shell structure with an $n=1$ and $n=2$ shell being present in the trap. In order to quantify the observed profiles, we fit parabola shapes according to a shell model with two (for small atom numbers) or three shells (for the profile corresponding to an atom number of $3.5\times 10^5$ atoms) to our data. The radii of the shells as well as the slice thickness $d$ are fitted independently.  We observe that with increasing atom number, the size of the inner $n=2$ Mott shell grows, until for larger atom numbers a third shell is formed in the center of the cloud, which we observe as an additional parabola on top of the truncated parabola of $n=1$ sites in Fig.~\ref{fig:profiles}d.

\begin{figure}
\includegraphics[scale=0.9]{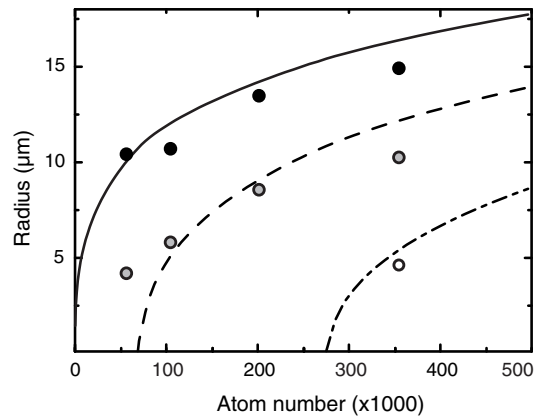}
\caption{Fitted radii of the Mott shells for different atom numbers and a lattice depth of $22~E_{\text{r}}$. The solid, grey and hollow data points denote the radii of the shells with single, double and triple occupation, respectively. The solid, dashed and dashed dotted lines are calculations of these first three Mott shell radii for zero temperature and zero tunnelling.} 
\label{fig:radii}
\end{figure}

The evolution of the sizes of the Mott shells when increasing the atom number is shown in Fig~\ref{fig:radii}. 
Here we plot the independently fitted shell radii versus the mean atom number of the cloud. 
The expected radii have been calculated for our experimental parameters assuming no tunneling and zero temperature \cite{demarco05a}. The distinct growth of successive shells with increasing atom number is in good agreement with theory, although our experimentally measured radii seem to be slightly lower than the expected values for larger atom numbers.

Direct observation of the density distribution also makes it possible to investigate the evolution of the shell structure during the superfluid to Mott insulator transition. For this, we compare the size of the distribution of doubly occupied sites to the size of the overall distribution, and therefore plot the ratio of their respective radii $R_{n=2}/R_{\text{total}}$.
Fig.~\ref{fig:radiiratio} shows the evolution from the superfluid state to the Mott insulating state for a fixed atom number of $N_{\text{total}} =(1.0\pm0.1)\times10^5$. The lattice depth was varied between $3~E_{\text{r}}$ and $40~E_{\text{r}}$. The corresponding trap frequencies of the external parabolic confinement range from 2$\pi\times$28~Hz to 2$\pi\times$80~Hz. To determine the radii, we fit a Thomas-Fermi distribution with free central density and radius to the respective profiles. 
This distribution is the theoretically predicted one in the weakly interacting limit, but was used here as a compromise for all profiles in order to obtain directly comparable values. When crossing the SF-MI transition, we observe a drop in the ratio $R_{n=2}/R_{\text{total}}$ from $\approx0.9$ below to $\approx0.6$ above the transition. A theory estimate shows that for our trap configuration, between the limits of a weakly interacting BEC and the limit of a MI, a drop from $\approx1$ to $\approx0.45$ is expected due to the pronounced redistribution of the atomic density within the trap. Our data suggests that most of the density redistribution happens already just before the Mott transition is reached at a lattice depth around $13\,E_{\text{r}}$ for our experimental parameters. 

\begin{figure}
\includegraphics[scale=0.9]{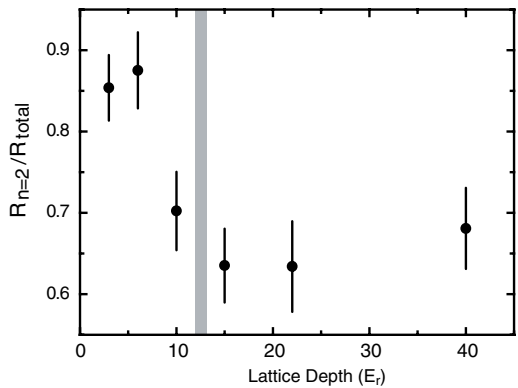}
\caption{Evolution of the shell structure through the superfluid to Mott-insulator transition for $N_{\text{total}}=(1.0\pm0.1)\times 10^5$ atoms.
Shown is the relative size of the area in which doubly occupied lattice sites are found compared to the total cloud size. The grey bar marks the lattice depth at which the SF to MI transition is expected to occur for a site occupation $n=1$.
} 
\label{fig:radiiratio}
\end{figure}


In conclusion, we have observed the density redistribution driven by the transition to an incompressible quantum gas, leading to a shell structure in the in-trap density distribution of the Mott insulator state.
As temperatures on the scale of the on-site interaction energy $U$ strongly affect the shape of such a shell structure \cite{demarco05a,gerbierinprep}, this method can provide an effective way to measure the temperature of the system inside the lattice potential. 
Due to the possibility of freezing out the momentary state within 100\,$\mu$s, the time-evolution of the density distribution can also be observed. This technique therefore could allow for studying the dynamics of excitations in the Mott insulator. The local manipulation of the spins within the trapped atom cloud with high spatial resolution also offers novel possibilities to create and observe spin excitations in the lattice system.

\begin{acknowledgments}
We acknowledge financial support from the DFG and the EU under a Marie Curie excellence grant (QUASICOMBS) and a Specific Targeted Research
Project (OLAQUI).
\end{acknowledgments}



\end{document}